\documentclass{article}
\usepackage[utf8]{inputenc}
\usepackage{authblk}
\usepackage[dvips]{graphicx}
\graphicspath{{noiseimages/}}
\usepackage{xcolor}
\usepackage[T2A]{fontenc}
\usepackage{tikz}
\usepackage{pgfplots}
\usepackage{mathrsfs}
\pgfplotsset{compat=1.7}
\usetikzlibrary{intersections, pgfplots.fillbetween}
\usetikzlibrary{decorations.pathmorphing}
\usetikzlibrary{arrows.meta}
\usepackage[mode=buildnew]{standalone}
\usepackage[toc,page]{appendix}
\usepackage{amsmath}
\usepackage{amssymb}
\usepackage{float}
\usepackage{comment}
\usepackage{a4wide}
\usepackage[english]{babel}
\usepackage{hyperref}
\definecolor{urlcolor}{HTML}{990000}
\definecolor{linkcolor}{HTML}{005F5F}
\hypersetup{pdfstartview=FitH,  linkcolor=linkcolor,urlcolor=urlcolor, colorlinks=true,citecolor=blue}
\usepackage[page]{appendix}

\author[1,2]{E.T. Akhmedov \footnote{\tt akhmedov@itep.ru} }
\author[1,3]{D.V. Diakonov\footnote{\tt dmitrii.dyakonov@phystech.edu}}

\affil[1]{Institutskii per. 9, Moscow Institute of Physics and Technology, 141700, Dolgoprudny, Russia}
\affil[2]{B. Cheremushkinskaya, 25, Institute for Theoretical and Experimental Physics, 117218, Moscow, Russia} 
\affil[3]{\itshape Bol'shoi Karetnyi per., 19, Institute for Information Transmission Problems, 127994, Moscow, Russia}

\title{\textcolor{black}{Sine-Gordon solitons in AdS, dS and other hyperbolic spaces}}

\begin{document}

\numberwithin{equation}{section}

\maketitle

\begin{abstract}
We find infinitely many soliton-like solutions in a deformation of the sine-Gordon theory in $(d+1)$-dimensional $AdS_{d+1}$ (anti-de Sitter) spacetime for $d \geq 2$, as well as single solitonic solutions in $dS_{d+1}$ (de Sitter) and $\mathrm{H}{d+1}$ (Lobachevsky) spaces for $d \geq 1$ and in $AdS_2$. We also find a deformation of the kink solution in scalar field theory with a polynomial potential in $AdS_2$. The deformation of the sine-Gordon theory strikingly resembles the bosonic part of the flat-space supersymmetric sine-Gordon theory. In the infinite radius limit, single soliton solutions reduce to solitons in flat space. Meanwhile, the multisoliton solution of $AdS{d+1}$, $d\geq 2$ for certain values of the parameters reduces in the same limit to a single soliton solution boosted in the normal direction. However, there are also multisoliton solutions in $AdS_{d+1}$, $d \geq 2$ that do not have a flat space limit.
\end{abstract}

\newpage

\tableofcontents

\section{Introduction}

Quantum field theory beyond tree-level in de Sitter ($dS$) \cite{Krotov:2010ma,Akhmedov:2013vka,Akhmedov:2024npw,Akhmedov:2019cfd,Akhmedov:2022uug,Akhmedov:2021rhq,Palma:2025oux,Miao:2024shs,Pimentel:2026kqc,Moreau:2020gib,Moreau:2019jpn}
and in anti-de Sitter ($AdS$) space-times
\cite{Akhmedov:2020jsi,Akhmedov:2018lkp,Akhmedov:2012hk,Melton:2025ecj,Bertan:2018khc} is very challenging. This difficulty calls for a simple model that can be exactly solved in these spaces, is not conformal (to distinguish these spaces from flat space), and does not rely on the large $N$ expansion (which may omit some important loop diagrams). In this respect, a promising example is given by the two-dimensional sine-Gordon theory. The question is: is it solvable in any hyperbolic space? If the answer is yes, then in what sense is it solvable?

At the classical level in flat space, the exact solvability of such a model can be seen through the presence of an infinite family of integrals of motion. However, in curved spaces there is no conservation in the proper flat-space sense even if it is present: covariant conservation of a quantity is supposedly not sufficient. Perhaps a hint of solvability can be seen via the presence of infinitely many soliton solutions in the model. The goal of our paper is to understand whether there exist soliton solutions in sine-Gordon theory in such hyperbolic spaces as $dS$, $AdS$, and Lobachevsky space.

Solitons are regular, stable, particle-like solutions that arise as exact solutions of nonlinear partial differential equations in flat space. There are of course also higher dimensional extentions of solitons. They appear not only in pure mathematics but also in various physical systems, where the balance between nonlinearity and dispersion prevents the wave packet from spreading \cite{10.1063/1.1654836,1973IEEEP..61.1443S}. There is a close connection between the existence of soliton solutions and the integrability of theories via the inverse scattering transform (see e.g. \cite{Gardner:1967wc}). The sine-Gordon model in two dimensions stands as a canonical example of such integrable theories \cite{Vergeles:1976ra,Zamolodchikov:1978xm,1972JPSJ...33.1459H}. In higher dimensions in flat spacetime, Derrick's theorem states that there exist no stable time-independent solutions of finite energy \cite{Derrick:1964ww}.
Extending the analysis of sine-Gordon solitons to higher-dimensional $AdS$ spacetimes is especially compelling because the confining nature of the $AdS$ gravitational potential enables stable localized configurations that would otherwise disperse in flat spacetime.

In this paper we consider the following deformation of the sine-Gordon theory:

\begin{align}
    \Box \phi \pm m^2 \sin \phi \pm \frac{2 d m}{R} \sin \frac{\phi}{2}=0, 
\end{align}
where the ``$+$'' sign corresponds to $dS_{d+1}$ space, while the ``$-$'' sign corresponds to $AdS_{d+1}$ and 
$\mathrm{H}_{d+1}$ spaces. Here $R$ is the radius of the corresponding symmetric hyperbolic space.

This theory reduces in the limit $m/R \to 0$ to the standard sine-Gordon theory, but in its own right strikingly resembles a supersymmetric theory in flat spacetime \cite{Inami:1995np}, with the action: 
\begin{align}
    S=\int d^2 x \left(\frac{1}{2}\partial_\mu \phi \partial^\mu \phi-i\bar{\psi} \gamma^\mu \partial_\mu \psi +m^2 \cos \left(\phi\right) +2 m \bar{\psi} \psi \cos\left( \frac{\phi}{2}\right) \right),
\end{align}
for a particular value of $\bar{\psi} \psi$.

We find the following soliton solutions in the theory under consideration over hyperbolic spaces:

\begin{itemize}
    \item In $(d+1)$-dimensional $AdS$ spacetime with $d \ge 2$, the infinite family of $N$-soliton solutions is as follows:
\begin{align}
    \phi= 4 \arctan \left[ (X \cdot \eta_1 )^{mR} F\left(\frac{(X \cdot\eta_{i_1}) }{(X \cdot\eta_{j_1})},...,\frac{(X \cdot\eta_{i_p}) }{(X \cdot\eta_{j_p})}\right) \right],
\end{align}
where $X$ are ambient spacetime coordinates and $\eta_i$, $i=\overline{1,N}$ is a set of mutually orthogonal null vectors from a two-dimensional null vector space, $(\eta_i \cdot \eta_i)=0$ and $(\eta_i \cdot \eta_j) = 0$ for $i \neq j$. This null vector space is present in the ambient spacetime. Furthermore, we require $m R \in \mathbb{N}$ to avoid complex arguments and complex $\phi$, since $X \cdot \eta$ does not have a definite sign, and $F(x_1, x_2, \dots, x_p)$ is a generic differentiable function.

\item In $(1+1)$-dimensional $AdS$, $(d+1)$-dimensional $dS$ spacetimes and in $(d+1)$-dimensional Lobachevsky ($\mathrm{H}_{d+1}$) space, we can construct only one solitonic solution:
\begin{align}
    \phi= 4 \arctan\left[  (X \cdot \xi )^{m R}\right],
\end{align}
where $mR\in \mathbb{N}$ in $AdS_{1+1}$ and $dS_{d+1}$, and $mR\in \mathbb{R}^+$ in $\mathrm{H}_{d+1}$. In all these cases there is only a single linearly independent light-like vector $(\xi \cdot \xi)=0$ in the ambient spacetime.

\end{itemize}

For the deformation of the $\phi^4$ potential:
\begin{align}
    \Box \phi - 2 m ^2 (\phi^2-1)\phi + \frac{m  d}{R} (\phi^2-1)=0,
\end{align}
we find solitonic solutions as follows: 
\begin{align}
    \phi = \tanh \left(\log\left[(X \cdot \eta_1 )^{mR} F\left(\frac{(X \cdot\eta_{i_1}) }{(X \cdot\eta_{j_1})},...,\frac{(X \cdot\eta_{i_p}) }{(X \cdot\eta_{j_p})}\right) \right] \right),
\end{align}
in $AdS_{d+1}$ for $d\geq 2$. Similarly in $(1+1)$--dimensional $AdS$ and in $\mathrm{H}_{d+1}$ for any $d$, there are only single soliton solutions. But in $dS_{d+1}$ due to the sign change of the potential the situation becomes unstable.

Understanding the stability of these solutions and whether these systems are integrable are important questions. We show that the static one-soliton solution in $AdS_{d+1}$ is stable under linear perturbations if: 
\begin{align}
    0 < m \le \frac{d-1}{2},
\end{align}
which means that since in our case $m$ must be an integer, stable solutions exist for $m\in\{1,2, \dots ,\left\lfloor\frac{d-1}{2}\right\rfloor\}$. However, this solution has infinite energy for $m<\frac{d+1}{2}$, which is a common situation in $AdS$ space due to the peculiar behavior of fields near the boundary of the spacetime.

\section{Geometry of hyperbolic spaces and hyperbolic plane waves}

A $(d+1)$-dimensional maximally symmetric space can be embedded in a flat spacetime of $d+2$ dimensions. $AdS$ spacetime is the hyperboloid embedded in a $(d+2)$-dimensional ambient flat spacetime with signature $(-,-,+,...,+)$:
\begin{align}
AdS_{d+1}=\{ X  \in \mathbf{R}^{2,d}, \ (X\cdot X)=X_\alpha X^\alpha =-R^2 \}, \quad \alpha = \overline{1,d+2}.
\end{align}
A $(d+1)$--dimensional $dS$ spacetime is the hyperboloid embedded in a $(d+2)$-dimensional ambient flat spacetime with signature $(-,+,+,...,+)$:
\begin{align}
dS_{d+1}=\{ X  \in \mathbf{R}^{1,d+1}, \ (X\cdot X)=X_\alpha X^\alpha =R^2 \}, \quad \alpha = \overline{1,d+2}.
\end{align}
A $(d+1)$--dimensional Lobachevsky space is the hyperboloid embedded in a $(d+2)$-dimensional ambient flat spacetime with signature $(-,+,+,...,+)$:
\begin{align}
\mathrm{H}_{d+1}=\{ X  \in \mathbf{R}^{1,d+1},X^0>0, \ (X\cdot X)=X_\alpha X^\alpha =-R^2 \}, \quad \alpha = \overline{1,d+2}.
\end{align}
In what follows, we will set the radii of these hyperboloids to $R = 1$, if not otherwise stated. 

In the ambient spacetime one can define null vectors $\xi$:
\begin{align}
    (\xi\cdot \xi)=0,
\end{align}
For a general signature\footnote{One can of course consider more general symmetric spaces via complexification. Namely, one can consider the following complex hyperboloid $Z_\alpha Z^\alpha = 1$, $Z_\alpha = X_\alpha + i Y_\alpha$, $\alpha = \overline{1,d+2}$. Taking various real sections of such a hyperboloid (i.e., setting some fraction of $X_\alpha$ and/or $Y_\alpha$ to $0$) one can obtain the sphere, $dS$, $AdS$, $\mathrm{H}$, and more generic symmetric spaces with $SO(p,q)$ generating group and corresponding stabilisers of generic points. One can straightforwardly generalise solitions found in our paper to these spaces.} $\mathbf{R}^{p,q}$, there are $\min(p,q)$ linearly independent orthogonal real null vectors. 
E.g., for $d>1$ in $AdS_{d+1}$ there exists a two-dimensional null vector space $V_\eta$. For example, for $d=2$ one can consider two linearly independent orthogonal real null vectors: $\xi_a=(1,0,0,1)$ and $\xi_b=(0,1,1,0)$: 
\begin{align}
\label{ort and null}
    (\xi_a \cdot \xi_a)=0, \quad  (\xi_b \cdot \xi_b)=0 \quad \text{and} \quad (\xi_a\cdot \xi_b)=0.
\end{align}
Hence any linear combination: 
\begin{align}
    \eta_i= a_i\xi_a+b_i \xi_b
\end{align}
is null and orthogonal to any other null vector from $V_\eta$: 
\begin{align}
  (\eta_i \cdot \eta_j)=0.
\end{align}
For $dS$ and Lobachevsky spaces, the ambient spacetime is given by $\mathbf{R}^{1,d+1}$. Hence, the dimension of the null vector space is one. 

Now let us introduce the following modes (we will refer to them as hyperbolic plane waves):
\begin{align}
    f_\lambda(X) = (X\cdot \xi)^\lambda
\end{align}
which solve the Klein-Gordon equation with an arbitrary parameter $\lambda$ that defines the mass of the field:
\begin{align}
   \Box (X \cdot \xi)^\lambda=-\frac{1}{X \cdot X}\lambda(\lambda+ d)  (X \cdot\xi)^\lambda
   + \lambda(2\lambda-1) (X \cdot\xi)^{\lambda-2} (\xi \cdot \xi),
\end{align}
if $X\cdot X = \pm 1$ and $\xi \cdot \xi = 0$: 
\begin{align}
\label{kg1}
   \Box (X \cdot \xi)^\lambda=-\frac{1}{X \cdot X}\lambda(\lambda+ d)  (X \cdot\xi)^\lambda
   .
\end{align}
This is a general form for any maximally symmetric space: for $dS$ \cite{Bros:1994dn,Bros:1995js}, $AdS$ \cite{Moschella:2025lqy}, Lobachevsky space \cite{Moschella:2007zza}, and for the sphere with complex-valued $\xi$ \cite{Akhmedova:2019bau}. To prove \eqref{kg1}, one can use the fact that covariant derivatives can be written in terms of projections of the derivatives in the ambient spacetime:
\begin{align}
    \nabla_\alpha =\left(\eta_{\alpha \beta}-\frac{X_\alpha X_\beta}{X\cdot X}\right) \partial^\beta.
\end{align}
Furthermore, in the limit $R\to \infty$ one can show that in $AdS$ spacetime this hyperbolic plane wave becomes the usual exponential: 
\begin{align}
    \lim_{R\to \infty} \left(\frac{X \cdot\xi}{R}\right)^{\lambda R} \sim e^{p_\mu x^\mu},
\end{align}
with the spacelike vector $p_\mu$: 
\begin{align}
    p^2 =-p_0^2+\vec{p}^2 =\lambda^2, 
\end{align}
whose components are expressed via components of the null vector $\xi$. To prove this fact one can consider the Poincaré coordinate parametrization of $AdS$, in which: 
\begin{align}
\label{Poincare}
\begin{cases}
X_{1}=R{\frac {z}{2}}\left(1+{\frac {1}{z^2}}\left(1+\frac{{x}^{2}-t^{2}}{R^2}\right)\right)
\\X_{2}={\frac {1}{z }}t
\\X_{i}={\frac {1}{z }}x_{i-2}  \quad \quad i \in (3,...d+1)
\\X_{d+2}=R{\frac { z}{2}}\left(1-\frac{1}{z^2}\left(1-\frac{{x}^{2}-t^{2}}{R^2}\right)\right)\end{cases} \quad \text{and} \quad \begin{cases}
\xi_{1}=1/R
\\ \xi_{2}=p_0
\\\xi_{i}= p_{i-2} \quad \quad i \in (3,...,d+1)
\\\xi_{d+2}=p_d
\end{cases}.
\end{align}
with $z=e^{x_d/R}$. 

Note that in $dS$ spacetime and Lobachevsky space the situation is different: one obtains a similar exponential with a timelike vector $p^2=-\lambda^2<0$ and a real-valued vector $p^2=\lambda^2>0$, respectively.  

The main properties of these hyperbolic plane waves follow from the fact that they obey the following relation: 
\begin{align}
\label{norm of tang der}
    \nabla_\mu (X \cdot\eta_i)\nabla^\mu (X \cdot\eta_j)= -\frac{1}{X\cdot X} (X \cdot\eta_i) (X \cdot\eta_j)+(\eta_i \cdot\eta_j).
\end{align}
Then, if the vectors $\eta_i$ and $\eta_j$ are orthogonal, i.e., they lie in the two-dimensional null space $\eta_i \in V_\eta$, we obtain the relation that is crucial in the construction of the infinite family of solitonic solutions in $AdS_{d+1}$:
\begin{align}
\label{main properties}
    \nabla_\mu (X \cdot\eta_i)\nabla^\mu (X \cdot\eta_j)=  (X \cdot\eta_i) (X \cdot\eta_j).
\end{align}
Furthermore, from this property one can find other identities that we will use below: 
\begin{align}
  \nabla^\mu (X \cdot\eta_k)   \nabla_\mu \frac{(X \cdot\eta_i) }{(X \cdot\eta_j)}=0 \quad \text{and} \quad  \nabla^\mu \frac{(X \cdot\eta_k) }{(X \cdot\eta_l)}   \nabla_\mu \frac{(X \cdot\eta_i) }{(X \cdot\eta_j)}=0
\end{align}
and: 
\begin{align}
   \Box\frac{(X \cdot\eta_i) }{(X \cdot\eta_j)}=0.
\end{align}
These observations lead to the following relations for any differentiable functions $F$ and $G$:
\begin{align}
\label{a1}
    \nabla^\mu G((X \cdot\eta_{k_1}),..., (X \cdot\eta_{k_n})) \nabla_\mu F\left(\frac{(X \cdot\eta_{i_1}) }{(X \cdot\eta_{j_1})},...,\frac{(X \cdot\eta_{i_p}) }{(X \cdot\eta_{j_p})}\right)=0.
\end{align}
\begin{align}
\label{a2}
    \nabla^\mu F\left(\frac{(X \cdot\eta_{i_1}) }{(X \cdot\eta_{j_1})},...,\frac{(X \cdot\eta_{i_p}) }{(X \cdot\eta_{j_p})}\right) \nabla_\mu F\left(\frac{(X \cdot\eta_{i_1}) }{(X \cdot\eta_{j_1})},...,\frac{(X \cdot\eta_{i_p}) }{(X \cdot\eta_{j_p})}\right)=0,
\end{align}
and
\begin{align}
\label{a3}
    \Box F\left(\frac{(X \cdot\eta_{i_1}) }{(X \cdot\eta_{j_1})},...,\frac{(X \cdot\eta_{i_p}) }{(X \cdot\eta_{j_p})}\right)=0.
\end{align}
We will use these identities below.

\section{Solitons in $AdS$ spacetime}

Motivated by the fact that the one-soliton solution of the sine-Gordon theory:
\begin{align}
    \Box \phi - m^2 \sin (\phi)=0,
\end{align}
has the following form: 
\begin{align}
    \phi=4 \arctan \left(e^{m \gamma \left(-v t+x\right)}\right)=4 \arctan \left(e^{ p_\mu x^\mu}\right),
\end{align}
where the vector $p_\mu$ is spacelike, i.e., $p^2=m^2$, and by the fact that in the flat-space limit the hyperbolic plane wave appears in the same form, i.e.,
\begin{align}
    \lim_{R\to \infty} (X \cdot \xi)^{m R} \sim e^{p_\mu x^\mu}, \quad \text{with} \quad p^2=m^2,
\end{align}
we can suppose that the one-soliton solution in $AdS$ spacetime should be of the form: 
\begin{align}
    \phi=4 \arctan \left((X^\alpha\xi_\alpha)^{m R}\right).
\end{align}
Nevertheless, it appears that this solves the double sine-Gordon equation:
\begin{align}
\label{double sin gor}
    \Box \phi -m^2 \sin \phi-2 d \frac{m}{R} \sin \frac{\phi}{2}=0,
\end{align}
where the second coupling constant depends on the dimension and radius of the ambient spacetime. In the flat space limit $R\to\infty$ one recovers the usual sine-Gordon theory.  
Note that the double sine-Gordon theory naturally arises in supersymmetric sine-Gordon theory \cite{Inami:1995np}, with action: 
\begin{align}
    S=\int d^2 x \left(\frac{1}{2}\partial_\mu \phi \partial^\mu \phi-i\bar{\psi} \gamma^\mu \partial_\mu \psi +m^2 \cos \left(\phi\right) +2 m \bar{\psi} \psi \cos\left( \frac{\phi}{2}\right) \right).
\end{align}

Furthermore, the energy of the field $\phi$, which we will use below, in Poincaré coordinates \eqref{Poincare} is given by: 
\begin{gather}
\label{energy}
    E= \nonumber\int d^{d-1}x \int_0^\infty \frac{dz }{z^{d}} \left( \frac{\partial_t \phi\partial_t \phi+\partial_z \phi\partial_z \phi+\partial_i \phi \partial_i\phi}{2}+\frac{m^2 \left(\cos \phi-1\right)+4 d m \left(\cos \frac{\phi}{2}-1\right)}{z^2}\right),
\end{gather}
and the minima and maxima of the potential terms depend on the dimension of space and the value of the parameter $m$. 
 
Note that the $n$-soliton solutions \cite{1972JPSJ...33.1459H} of the standard flat-space sine-Gordon theory: 
\begin{align}
\label{N-sol in sin g}
    \phi=4 \arctan \left(\frac{\sum_{n} A_ne^{p^\mu_n x_\mu}}{1+\sum_{k} B_k e^{p^\nu_k x_\nu}}\right).
\end{align}
cannot be generalized simply by replacing all exponents with the corresponding hyperbolic plane waves introduced above. Nevertheless, we can find other seemingly multiple soliton solutions in $AdS$ spacetime of the double sine-Gordon equation that do not reduce to \eqref{N-sol in sin g} in the flat-space limit. 

To find other solutions of \eqref{double sin gor} in the $AdS$ metric, let us use the ansatz:
\begin{align}
    \phi =4  \arctan \left(G F \right).
\end{align}
Then the double sine-Gordon equation can be rewritten in the form (with $R=1$): 
\begin{gather}
  F  \left( \Box-m (m+d) \right) G  
  +\\+\nonumber
  F^3 G \left(G \Box G-2 \nabla_\mu G \nabla^\mu G +m(m-d)G^2  \right)
  +\\+ \nonumber
  G\left(1+F^2 G^2\right) \Box F
  -
  2 F G^3 \nabla_\mu F \nabla^\mu F
  +
  2 \left(F^2G^2-1\right) \nabla_\mu G \nabla^\mu F
  =0.
\end{gather}
We can decompose this equation into three simpler equations: 
\begin{align}
\label{b1}
    \left[ \Box -m (m+d) \right] G=0,
\end{align}
\begin{align}
\label{b2}
    G \Box G-2 \nabla_\mu G \nabla^\mu G +m(m-d)G^2=0
\end{align}
and 
\begin{align}
\label{b3}
  2 \left(F^2G^2-1\right) \nabla_\mu G \nabla^\mu F
  -
  2 F G^3 \nabla_\mu F \nabla^\mu F
  +
   G\left(F^2 G^2+1\right) \Box F=0,
\end{align}
which perhups cover only part of the full set of solutions. The first equation \eqref{b1} coincides with the Klein-Gordon equation \eqref{kg1} for hyperbolic plane waves. Hence, the general solution should be a linear combination of plane waves: 
\begin{align}
\label{anz}
    G= \sum_i (X \cdot \xi_i )^m,
\end{align}
where $\xi_i$ are null vectors. 

To satisfy the second equation \eqref{b2}, one should impose the condition that this set of vectors $\xi_i$ are orthogonal to each other, i.e., $\xi \in V_\eta$. Indeed, using \eqref{kg1} and \eqref{main properties}, one can show that it is solvable using the ansatz: 
\begin{align}
\label{anzG}
    G= \sum_i (X \cdot \eta_i )^m,
\end{align}
where $\eta_i \in V_\eta$. 

As we can see, the terms in the third equation \eqref{b3} vanish identically if the function $F$ is defined as follows: 
\begin{align}
\label{anz F}
     F=F\left(\frac{(X \cdot\eta_{i_1}) }{(X \cdot\eta_{j_1})},...,\frac{(X \cdot\eta_{i_p}) }{(X \cdot\eta_{j_p})}\right),
\end{align}
due to the identities \eqref{a1}, \eqref{a2} and \eqref{a3}. As a result, a general solution is given by: 
\begin{align}
    \phi= 4 \arctan \left[\left( \sum_i (X \cdot \eta_i )^m\right) F\left(\frac{(X \cdot\eta_{i_1}) }{(X \cdot\eta_{j_1})},...,\frac{(X \cdot\eta_{i_p}) }{(X \cdot\eta_{j_p})}\right) \right],
\end{align}
where $m \in \mathbb{N}$, to avoid complex arguments since $X \cdot \eta$ does not have a definite sign.  Since: 
\begin{align}
    \sum_i (X \cdot \eta_i )^m= (X \cdot \eta_j )^m \left(   \sum_i\frac{(X \cdot \eta_i )^m}{(X \cdot \eta_j )^m}\right),
\end{align}
where the bracket contains a function of the ratio of hyperbolic plane waves, we can include it in the definition of $F$. Hence, the generic form of the solution can be written as: 
\begin{align}
\label{general sol}
    \phi= 4 \arctan \left[  (X \cdot \eta_1 )^m  F\left(\frac{(X \cdot\eta_{i_1}) }{(X \cdot\eta_{j_1})},...,\frac{(X \cdot\eta_{i_p}) }{(X \cdot\eta_{j_p})}\right) \right],
\end{align}
In the flat-space limit the parameter $m$ becomes continuous since if we restore the radius of $AdS$, we obtain $m=\frac{n}{R}$. One can take the flat space limit $R\to \infty$ of this general solution. Namely, in the flat space limit one obtains a form of the soliton in which all hyperbolic plane waves are replaced by the corresponding exponentials. As an example consider:
\begin{align}
\begin{cases}
X_{1}=R{\frac {z}{2}}\left(1+{\frac {1}{z^2}}\left(1+\frac{x^2}{R^2} -\frac{t^2}{R^2}\right)\right)
\\X_{2}={\frac {1}{z }}t
\\X_{3}={\frac {1}{z }}x 
\\X_{4}=R{\frac { z}{2}}\left(1-\frac{1}{z^2}\left(1-\frac{x^2}{R^2} +\frac{t^2}{R^2}\right)\right)\end{cases} \quad \text{and} \quad 
\eta_j=(-1,\omega_j/m,\omega_j/m,1)
\end{align}
Then in the flat space limit: 
\begin{align}
\left(\frac{X\cdot \eta_j}{R}\right)^{m R}=e^{-\omega_j t+ \omega_j x + m y}.
\end{align}
Note also that we take the limit for some particular form of the null vectors $\eta_j$. For example, for: 
\begin{align}
\eta=(0,a,a,0),
\end{align}
there is no well-defined flat space limit:
\begin{align}
\lim_{R \to \infty}\left(a\frac{x-t}{R}\right)^{m R}.
\end{align}

Then, using the fact that the function $F$ depends only on the ratio of the hyperbolic plane waves, we obtain: 
\begin{align}
    F\left(e^{-(\omega_{i_1}-\omega_{j_1})( t- x) },...,e^{-(\omega_{i_p}-\omega_{j_p})( t- x) }\right) = f(t-x).
\end{align}
Hence in the limit the solution reduces to: 
\begin{align}
\label{general sol M}
    \phi= 4 \arctan \left[ e^{-\omega (t-x)+ m y} f(t-x) \right]
    = 4 \arctan \left[  e^{ m y} \bar{f}(t-x) \right].
\end{align}
This is a sort of a wave packet of domain walls in the $y$ direction that moves with the speed of light along the $x$ direction. Thus, this flat space soliton solves separately the following two equations:
\begin{align}
    \phi''_{yy} = -\frac{\partial V(\phi)}{\partial \phi}, \quad \ddot{\phi} - \phi''_{xx} = 0.
\end{align}
Hence, the solution \eqref{general sol} for the particular choice of parameters describes a single soliton.

Finally, let us find the behavior of the soliton field in the vicinity of the boundary of $AdS$ spacetime, i.e., as $z \to 0$. In this limit $X \approx \frac{1}{z} X^b$, where:
\begin{align}
X^b=\begin{cases}
X^b_{1}= \frac{1}{z} R{\frac {1}{2}}\left(1+\frac{\vec{x}^2}{R^2} -\frac{t^2}{R^2}\right)
\\X^b_{2}={\frac {1}{z }}t
\\X^b_{i}={\frac {1}{z }}x^i 
\\X^b_{d+2}=-\frac{1}{z} R{\frac { 1}{2}}\left(1-\frac{\vec{x}^2}{R^2} +\frac{t^2}{R^2}\right)\end{cases}.
\end{align}
Then hyperbolic plane waves diverge in this limit, but their ratio is finite. In fact: 
\begin{align}
    \lim_{z\to 0}(X\cdot \xi)^m \approx\frac{1}{z^m} (X^b\cdot \xi)^m \to \infty  \times \operatorname{sign} (X\cdot \xi)^m. 
\end{align}
As a result, near the boundary the general solution \eqref{general sol} behaves as:
\begin{align}
    \phi \approx 2\pi  \times \operatorname{sign} \left( \phi\right).
\end{align}
This behaviour will be used below to estimate the energy of the solitions in $AdS$ spacetime.

\subsection{Polynomial potential}

Another potential for which one can find similar solitonic solutions in $AdS$ spacetime is as follows: 
\begin{align}
    \Box \phi -2 m ^2 (\phi^2-1)\phi +m  d (\phi^2-1)=0.
\end{align}
Using the ansatz: 
\begin{align}
    \phi = \tanh \left[\log\left(G F\right) \right],
\end{align}
we find the following equations: 
\begin{gather}
     F^2 G\left(1+G^2\right) \left(\Box G-m(m+d) G\right)
     +\\+ \nonumber
     F^2 \left(1-3 G^2\right) \left(\nabla_\mu G \nabla^\mu G-m^2 G^2\right)+\\+ \nonumber
     4 G F\left(1-F^2 G^2\right) \nabla_\mu G \nabla^\mu F+ G^2\left(1-3 G^2 F^2\right)  \nabla_\mu F \nabla^\mu F+ G^2 F\left(1+G^2 F^2 \right) \Box F=0.
\end{gather}
Then the first two lines vanish if $G$ is given by \eqref{anzG}, and the terms in the third line vanish separately if $F$ is given by \eqref{anz F}.

\subsection{Visualizing single soliton solution in $AdS_{1+1}$}

\begin{figure}[H]
    \centering
      \includegraphics[scale=0.09]{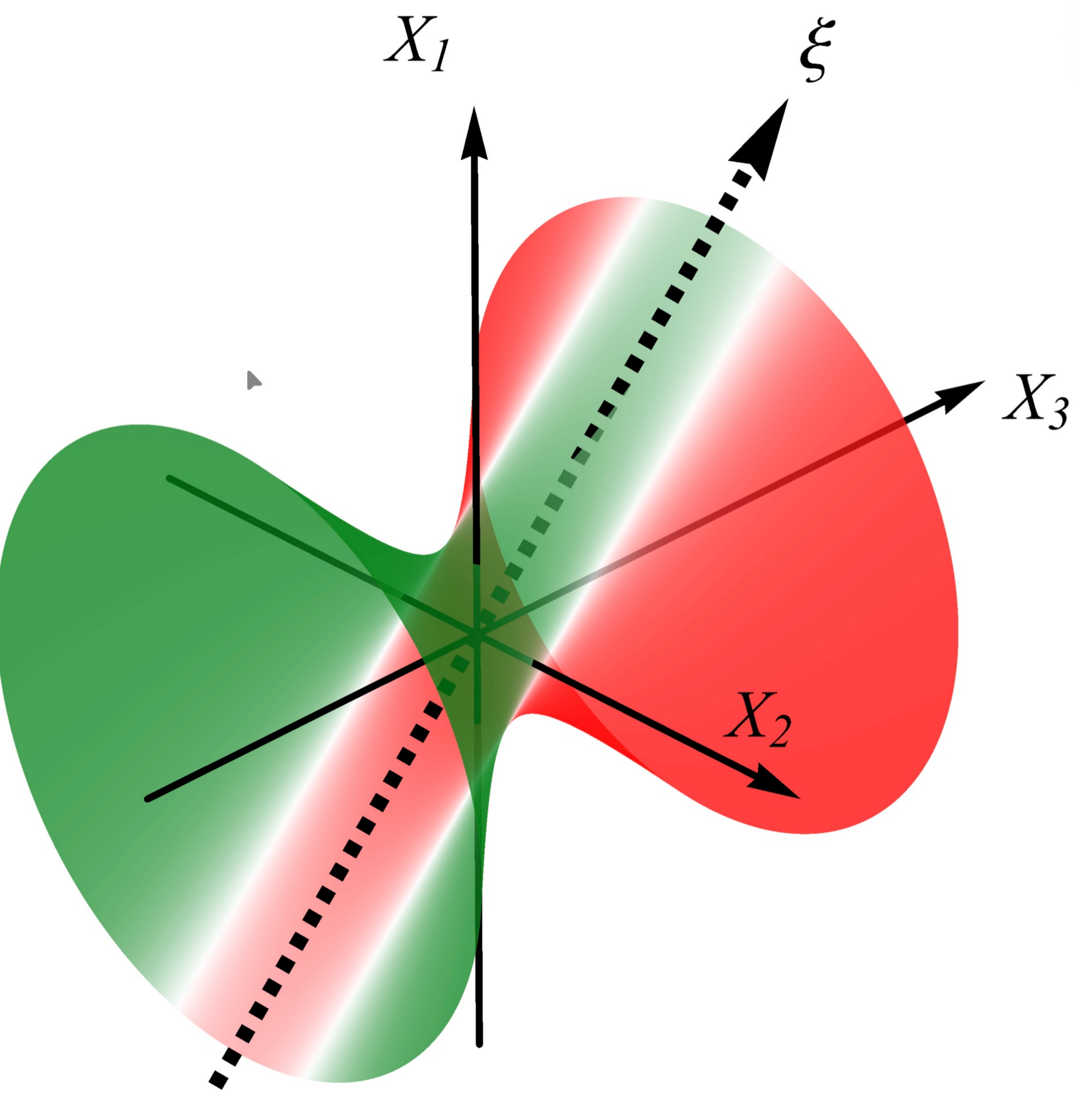} \quad \quad
  \includegraphics[scale=0.46]{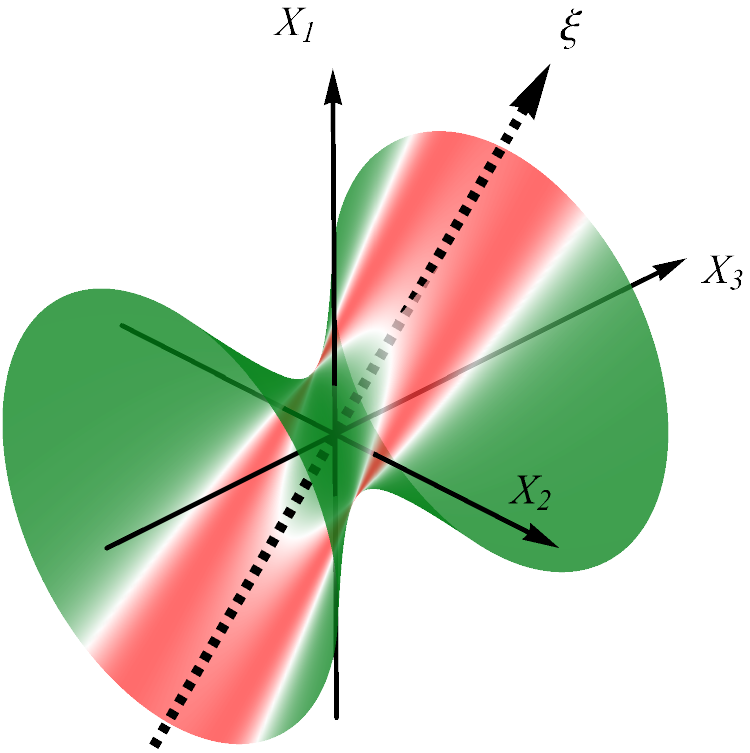} 
    \caption{The value of the $\phi$ field of the soliton in $AdS$ for $m=1$ (left) and $m=2$ (right) with the same null vector $\xi=(1,0,1)$ is shown. On the left picture, the field value changes from $0$ (red) to $2\pi$ (green). On the right picture, the field value changes from $-2\pi$ (red) to $2\pi$ (green). }
    \label{figAds2}
\end{figure}

Let us consider $1+1$ AdS space with signature $(-,-,+)$ in embedding coordinates. In this space we can construct only one solitonic solution:
\begin{align}
    \phi= 4 \arctan \left[(X \cdot \xi )^m \right],
\end{align}
since there are no two linearly independent null vectors. 
The graph of this solution is shown in Figure \ref{figAds2}, where we plot the value of the field on $AdS$ in the ambient coordinates; the dotted line shows the direction of the null vector $\xi$. 


\subsection{Visualizing multiple soliton solutions in $AdS_{2+1}$}

\begin{figure}[H]
    \centering
    \includegraphics[scale=0.27]{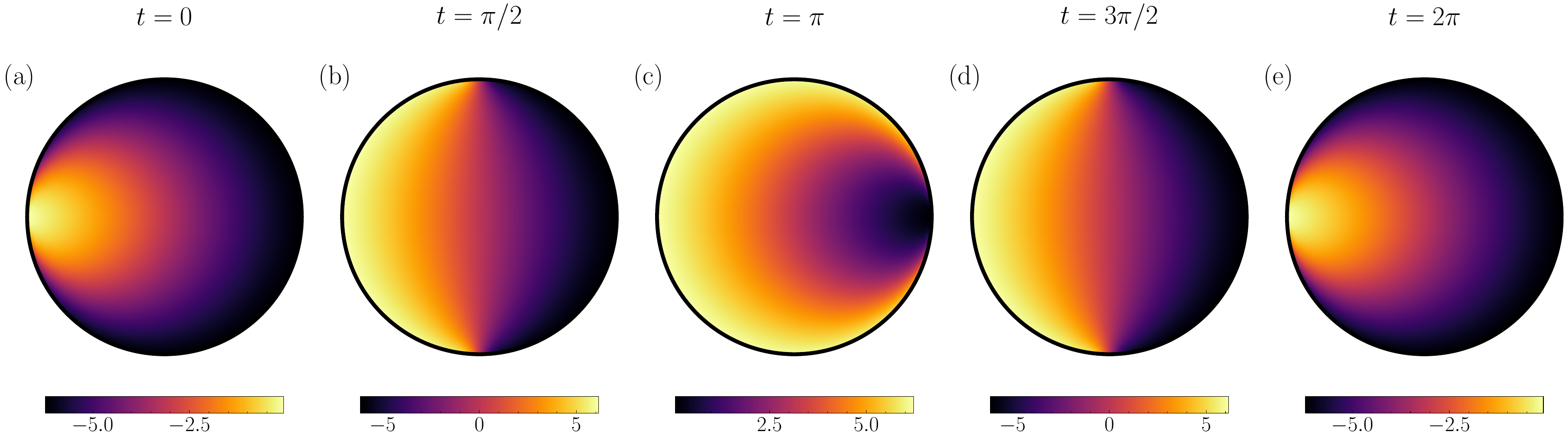}
    \caption{A set of graphs of $\phi$ in $AdS_{1+2}$ shown on the Poincaré disk for different moments in time $\tau\in(0,2\pi)$. The value of the $\phi$ field of the soliton for $m=1$ and null vector $\eta=(0,1,1,0)$. The value of the field changes from $-2\pi$ (purple) to $2\pi$ (yellow).   }
    \label{figAds3}
\end{figure}

Let us consider $2+1$ $AdS$ spacetime with the following embedding coordinates and null vectors: 
\begin{align}
\begin{cases}
X_{1}=\cosh(\rho) \cos (t)
\\X_{2}=\cosh(\rho) \sin (t)
\\X_{3}=\sinh(\rho) \cos (\theta) 
\\X_{4}=\sinh(\rho) \sin (\theta)
\end{cases}, 
\quad \quad \eta =(0,1,1,0)
\end{align}
The metric in these coordinates is given by: 
\begin{align}
    ds^2 =-\cosh ^2 \rho \, dt^2 +d\rho^2 +\sinh^2 \rho \, d\theta^2, 
 \end{align}
The constant time slices of this coordinate patch are Lobachevsky spaces: 
 \begin{align}
     ds^2_{t=\text{const}}= d \rho^2+\sinh^2\rho \, d\theta^2,
 \end{align}
which can be mapped to the Poincaré disc via the coordinate change: 
\begin{align}
x=\tanh \left(\frac{\rho}{2}\right) \cos(\theta), \quad 
y=\tanh \left(\frac{\rho}{2}\right) \sin(\theta).
\end{align}
In this space we can construct the simplest soliton solution: 
\begin{align}
    \phi_n= 4 \arctan \left[\left(X \cdot \eta  \right)^m\right].
\end{align}
We plot $\phi$ in global $AdS_{2+1}$ coordinates, shown on Poincaré discs for different moments of time $t\in(0,2\pi)$, in Figure \ref{figAds3}. There one can see how solitons move from one boundary to another, then reflect and move back. 



\subsection{Energy and Stability}

\begin{figure}[H]
    \centering
    \includegraphics[scale=0.3]{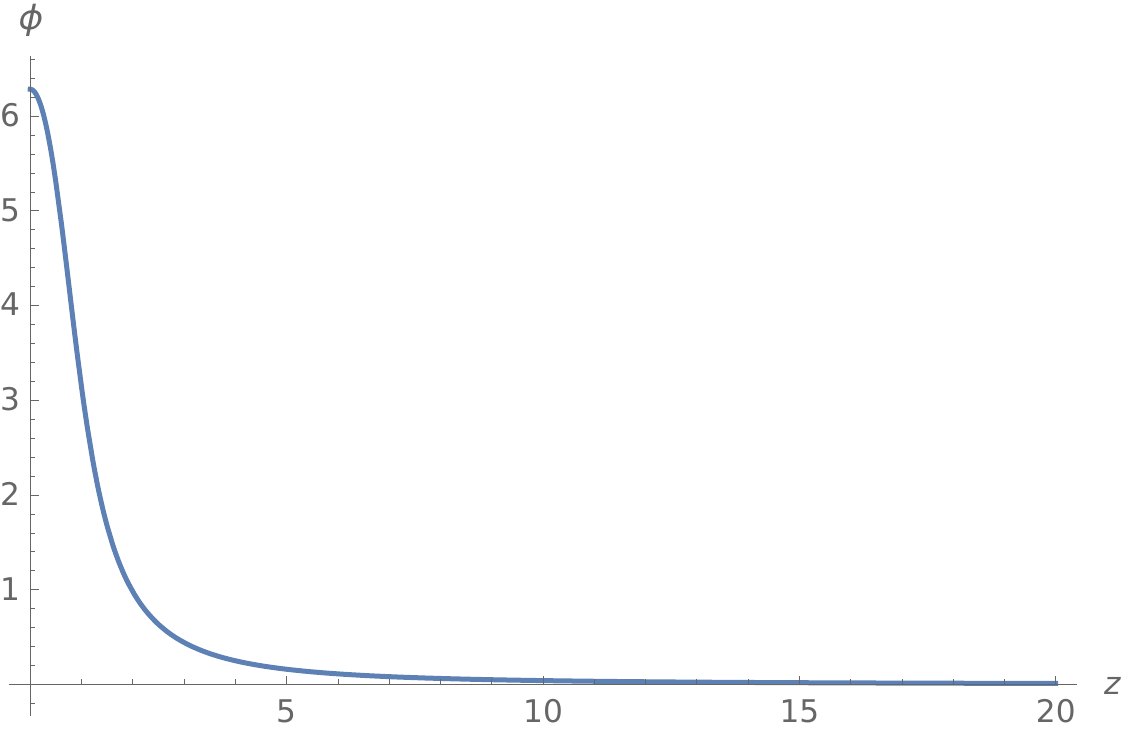}
    \caption{The value of the $\phi$ field in Poincaré coordinates of $AdS$ for $m=1$ with the null vector $\xi=(-1,0,-1)$.   }
    \label{AdsPoincare static}
\end{figure}

We want to verify the stability of the obtained solitons under linearized perturbations\footnote{It seems that at least depicted above two-diemnsional solitions in $AdS_2$ have some sort of topological stability, but that demands a more carefull study.}. Let us consider a one-soliton solution in $AdS_{d+1}$ in Poincaré coordinates \eqref{Poincare}. In these coordinates, the metric is:
\begin{align}
     ds^{2}=\frac {1}{z^{2}}\left(-dt^2+dz^{2}+d\vec{x}^2\right).
\end{align}
For simplicity, we choose a null vector:
\begin{align}
\xi=(-1,0,...0,-1),
\end{align}
such that the soliton is stationary:
\begin{align}
\label{solution}
    \phi_0=4  \arctan \left( z^{-m}\right).
\end{align}
We plot the graphic of this solution of Figure \ref{AdsPoincare static}

This static configuration might have infinite energy since contributions to the energy may diverge as $z \to 0$: 
\begin{align}
    E\sim \int dz \left(  8 d m^2 z^{2m-d-2}+... \right).
\end{align}
Thus, for $m\le \frac{d+1}{2}$, which is a quite common situation in $AdS$ spacetime, the classical solution has infinite energy due to the peculiar behavior of the field near the boundary. 

\begin{figure}[H]
    \centering
    \includegraphics[scale=0.3]{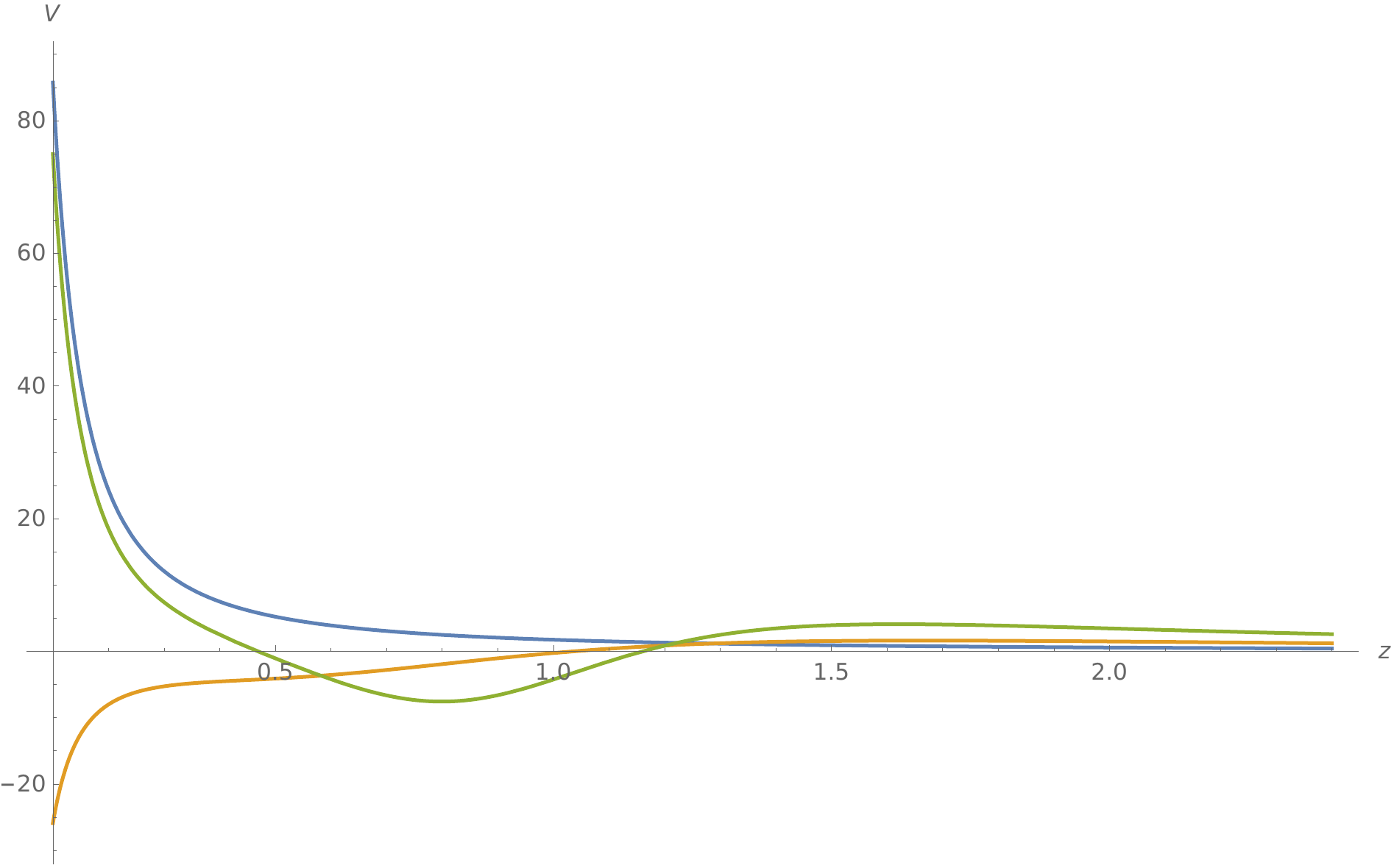}  
    \caption{The graph of the potential $V(z)$. Blue line: $0 < m \le \frac{d-1}{2}$, orange line: $\frac{d-1}{2} < m < \frac{d+1}{2}$, and green line: $m \ge \frac{d+1}{2}$.   }
    \label{v}
\end{figure}

Now we consider a small perturbation:
\begin{align}
    \phi=\phi_0+f,
\end{align}
and expand the equation to linear order in $f$.  Taking the ansatz $f=e^{-i \omega t+i p x} z^{\frac{d-1}{2}} \psi(z)$, we obtain a Schrödinger-type equation:
\begin{align}
    -\partial^2_z \psi(z)+V(z) \psi =\omega^2 \psi,
\end{align}
where the potential is:
\begin{align}
\label{pot}
    V(z)= \frac{1}{4 z^2}\left(2\left(\frac{4m}{1+z^{2 m}}-\frac{4m+d}{2} \right)^2-\frac{8m^2-d^2+2}{2}\right).
\end{align}
As $z \to 0$, it simplifies to:
\begin{align}
    V(z) \approx  \frac{1}{z^2}\left(m-\frac{d-1}{2}\right)\left(m-\frac{d+1}{2}\right),
\end{align}
thus, if $0< m \le \frac{d-1}{2}$, the potential \eqref{pot} is strongly positive and smoothly decreases to zero as $z \to \infty$; if $\frac{d-1}{2}< m < \frac{d+1}{2}$, the potential tends to minus infinity, hence the theory is unstable; in the case $m \ge \frac{d+1}{2}$, as $z \to 0$ the potential tends to infinity, but at some $z > 0$ it becomes negative, and therefore the theory is unstable.We plot this three cases on the Figure \ref{v}. 
As a result, stable solutions exist only for the case:
\begin{align}
    0< m \le \frac{d-1}{2}.
\end{align}
Hence, since in our case $m$ is an integer, stable under linearized perturbations solutions exist for $m\in\{1,2,\dots,\left\lfloor\frac{d-1}{2}\right\rfloor\}$.

\section{Soliton in $dS$ spacetime}

A $(d+1)$-dimensional $dS$ spacetime is the hyperboloid embedded into $(d+2)$-dimensional flat spacetime with signature $(-,+,...,+)$: 
\begin{align}
    X \cdot X= R^2.
\end{align}
Here, unlike the case of $AdS_{d+1}$ with $d\geq 2$, there is only one linearly independent null vector $\xi$. Furthermore, the main relation for the hyperbolic plane waves, which we have been using above, gives a different sign on the right-hand side of \eqref{kg} and \eqref{main properties} compared to the $AdS$ case. Namely:

\begin{align}
\label{kg}
   \Box (X \cdot \xi)^\lambda=-\lambda(\lambda+ d)  (X \cdot\xi)^\lambda
\end{align}
and 
\begin{align}
    \nabla_\mu (X \cdot\xi_i)\nabla^\mu (X \cdot\xi_j)= -(X \cdot\xi_i) (X \cdot\xi_j).
\end{align}
Due to this, we should change the sign of the potential terms in the double sine-Gordon equation: 
\begin{align}
    \Box \phi +m^2 \sin \phi+2 d \frac{m}{R} \sin \frac{\phi}{2}=0,
\end{align}
which is not a problem for a periodic potential, but does make the situation with the polynomial potential unstable. Hence, in $dS$ spacetime we can discuss only a modification of the sine-Gordon theory. 

Moreover, in the limit $R\to \infty$ the hyperbolic plane wave in $dS$ spacetime takes the usual exponential form: 
\begin{align}
    \lim_{R\to \infty} \left(\frac{X \cdot\xi}{R}\right)^{\lambda R} \sim e^{p_\mu x^\mu},
\end{align}
but with a timelike vector $p^2=-\lambda^2<0$.
This means that we cannot obtain a static configuration in the flat-space limit. The same happens with the usual sine-Gordon theory if we consider the potential with a minus sign:
\begin{align}
    \Box \phi +m^2 \sin \phi=0,
\end{align}
then the solution depends on time: 
\begin{align}
    \phi =4 \arctan \left( e^{p_\mu x^\mu}\right), 
\end{align}
such that in some frame we obtain: 
\begin{align}
    \phi =4 \arctan \left(e^{-p_0  t}\right), 
\end{align}
which describes a process --- the global change of the field from one extremum of the potential $\phi=2\pi$ at $t=-\infty$ to another extremum $\phi=0$ at $t=\infty$. 

If we apply the same method as in the previous section, we find only a one-soliton solution in any dimension: 

\begin{align}
    \phi =4 \arctan \left[ (X \cdot \xi)^m\right], 
\end{align}
where $m \in \mathbb{N}$ to avoid complex arguments since $X \cdot \xi $ does not have a definite sign. Note also that there is a complex null vector that is linearly independent from $\xi$, and we can construct infinitely many soliton solutions for a complex field $\phi$ with the same expression as \eqref{general sol}. 

\subsection{Visualizing the soliton in $dS_{1+1}$ spacetime}

\begin{figure}[H]
    \centering
    \includegraphics[scale=0.48]{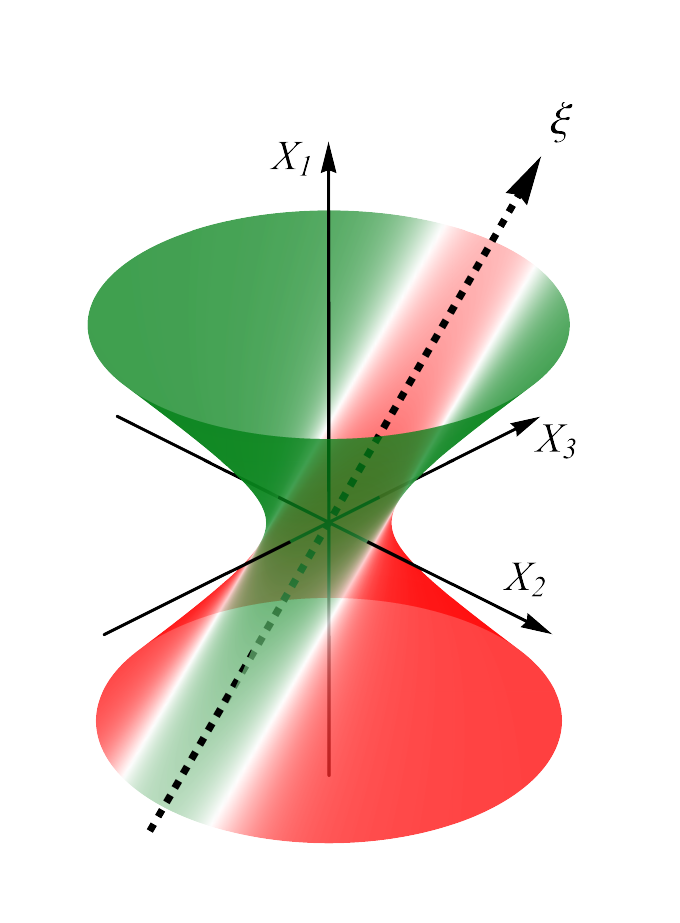}
       \includegraphics[scale=0.48]{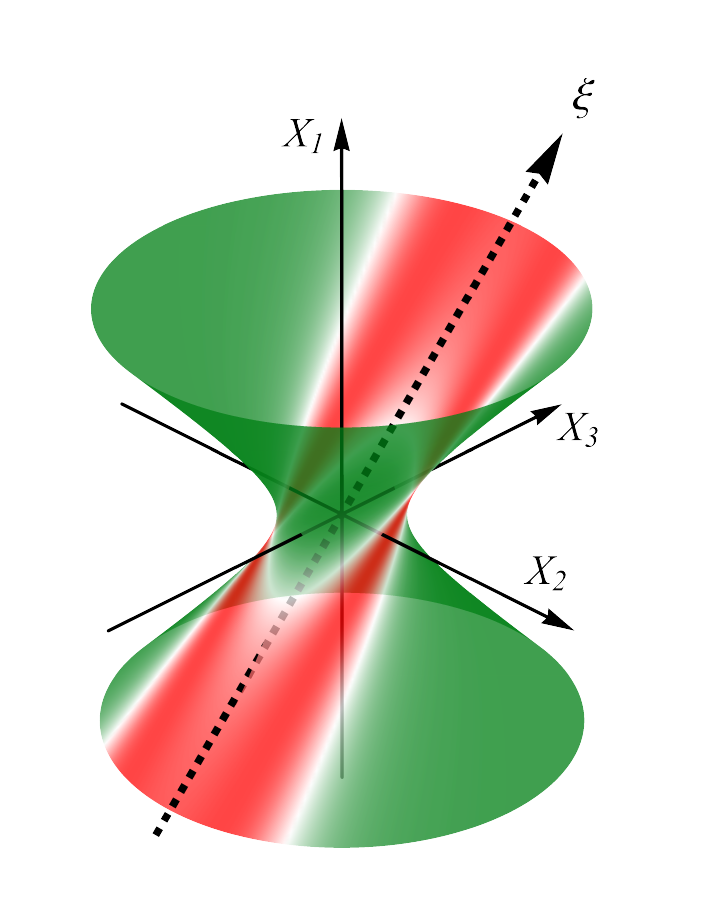}
    \caption{ The value of the $\phi$ field of the soliton in $AdS$ for $m=1$ and $m=2$ with the same null vector $\xi=(1,0,1)$. On the left picture, the field value changes from $-2\pi$ (red) to $2\pi$ (green). On the right picture, the field value changes from $0$ (red) to $2\pi$ (green). }
    \label{Glob de sitter 1}
\end{figure}

\begin{figure}[H]
    \centering
    \includegraphics[scale=0.3]{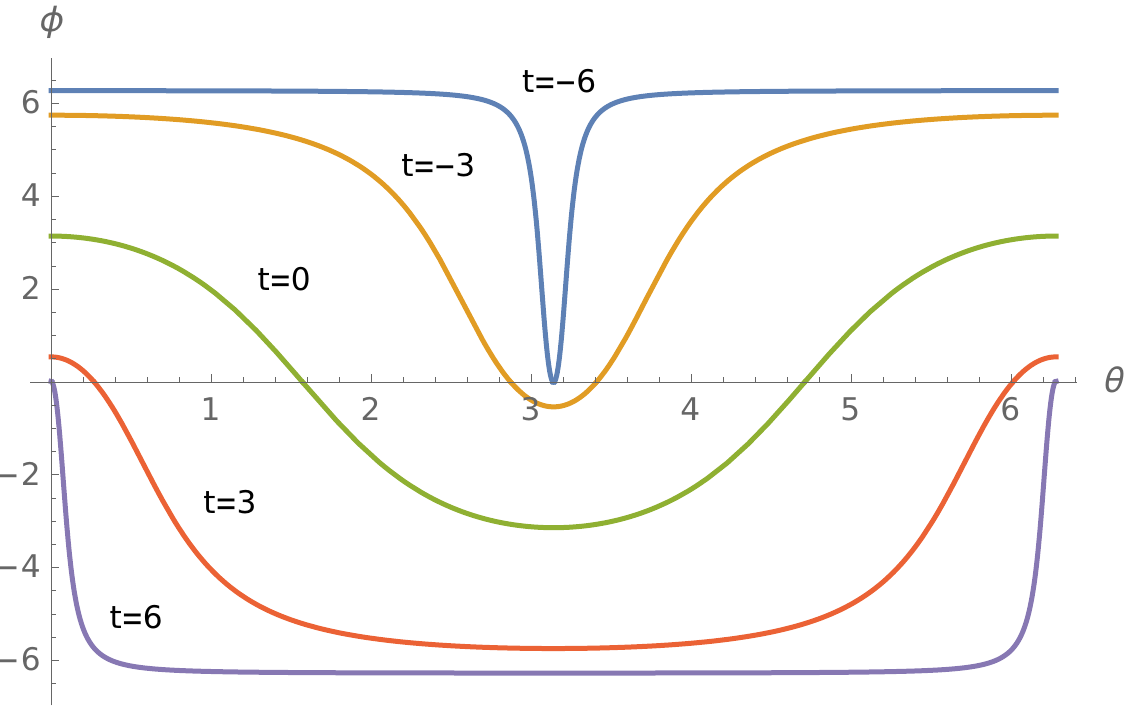}  
    \caption{The value of the $\phi$ field of the soliton in global coordinates of $dS$ for $m=1$ with the null vector $\xi=(1,1,0)$ for different moments in time.   }
    \label{Ds Glob m=1}
\end{figure}

\begin{figure}[H]
    \centering
    \includegraphics[scale=0.3]{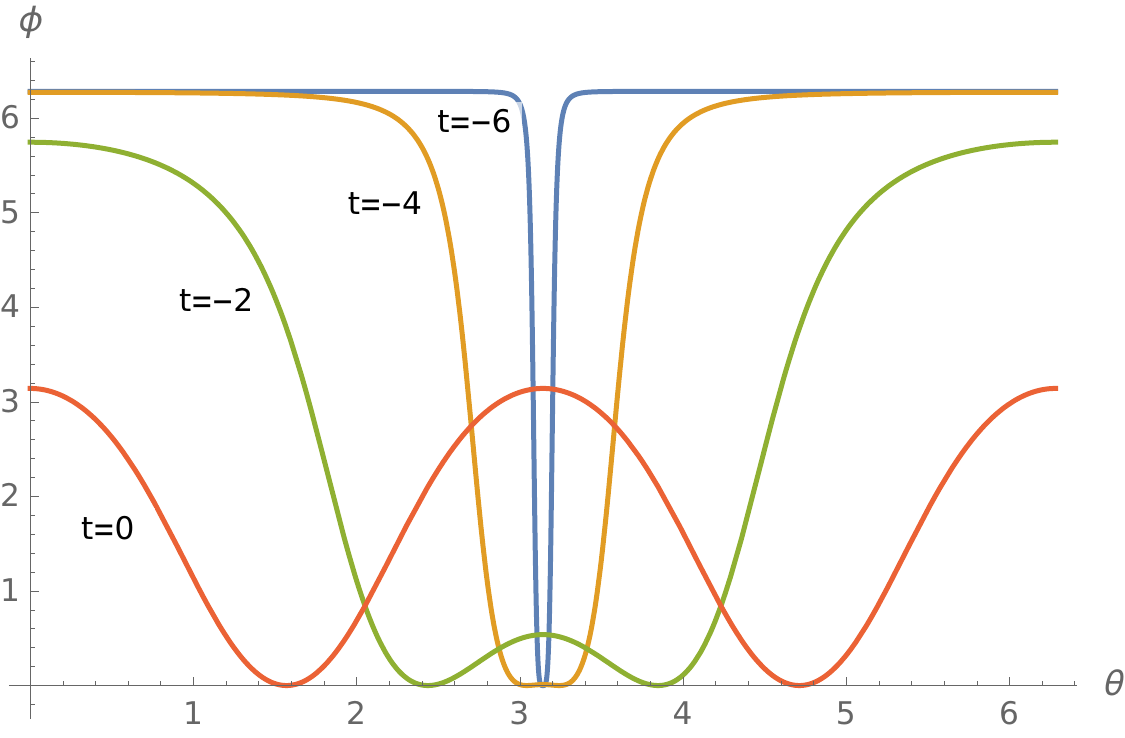}  
    \includegraphics[scale=0.3]{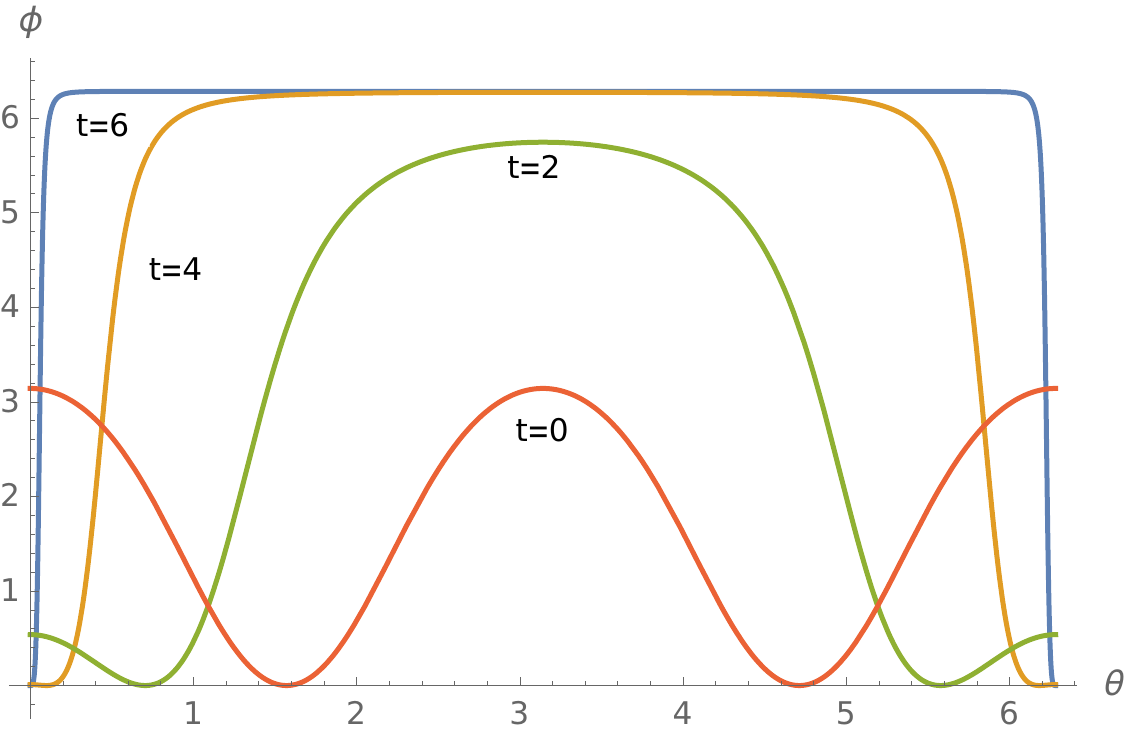}  
    \caption{The value of the $\phi$ field of the soliton in global coordinates of $dS$ for $m=2$ with the null vector $\xi=(1,1,0)$ for different moments in time.   }
    \label{Ds Glob m=2}
\end{figure}

Let us depict the soliton solution in two-dimensional $dS$ spacetime in global coordinates. In this case the embedding coordinates and null vector are given by: 
\begin{align}
\begin{cases}
X_{0}=\sinh(t)
\\X_{1}=\cosh(t)\cos(\theta)
\\X_{2}=\cosh(t) \sin(\theta)
\end{cases}
\quad \text{and} \quad
\begin{cases}
\xi_{0}=q
\\\xi_{1}=q\cos(\varphi)
\\\xi_{2}=q \sin(\varphi)
\end{cases}.
\end{align}
where we set the $dS$ radius to one, $R=1$. In these coordinates we have: 
\begin{align}
 X\cdot\xi =q\left(\cosh(t) \cos(\theta+\varphi)-\sinh(t) \right),
\end{align}
and as one can see, for any time there are points $\theta$ for which $X\cdot \xi=0$, i.e., where $\phi=0$.  In Figure \ref{Glob de sitter 1} we plot in embeding space how the one-soliton field depends on time for $m=1$ and $m=2$. In Figure \ref{Ds Glob m=1} and \ref{Ds Glob m=2} we plot in global coordinate now this configuration evolve in time. As one can see, the field interpolates from one extremum configuration to another, with one point at past and future infinity where the field vanishes:  
\begin{align}
    \phi =4 \arctan \left( X \cdot \xi(0)\right)  \quad \text{and} \quad    \phi =4 \arctan \left( (X \cdot \xi(0))^2\right).
\end{align}


As it should be figures in $dS_{1+1}$ and $AdS_{1+1}$ are just rotations of each other in the ambient spacetime.

\section{Soliton in Lobachevsky space}

Lobachevsky space is a hyperboloid embedded in flat spacetime with signature $(-,+,...,+)$:
\begin{align}
    X \cdot X= -R^2 \quad \text{and} \quad  X^0>0.
\end{align}
Here the double sine-Gordon equation is the same as for $AdS$ spacetime \eqref{double sin gor}. However, since there is only a one-dimensional null vector space, we obtain only a one-soliton solution:
\begin{align}
    \phi =4 \arctan \left[ (-X \cdot \xi)^m\right].
\end{align}
In this case one can show that for a future-oriented vector $\xi$, the value of $(-X\cdot \xi)$ is strictly positive. Because of this, there is no problem with the complex argument of the $\arctan$ function, and we can choose any value of $m$, i.e., $m\in \mathbb{R}$.

\begin{figure}[H]
    \centering
    \includegraphics[scale=0.5]{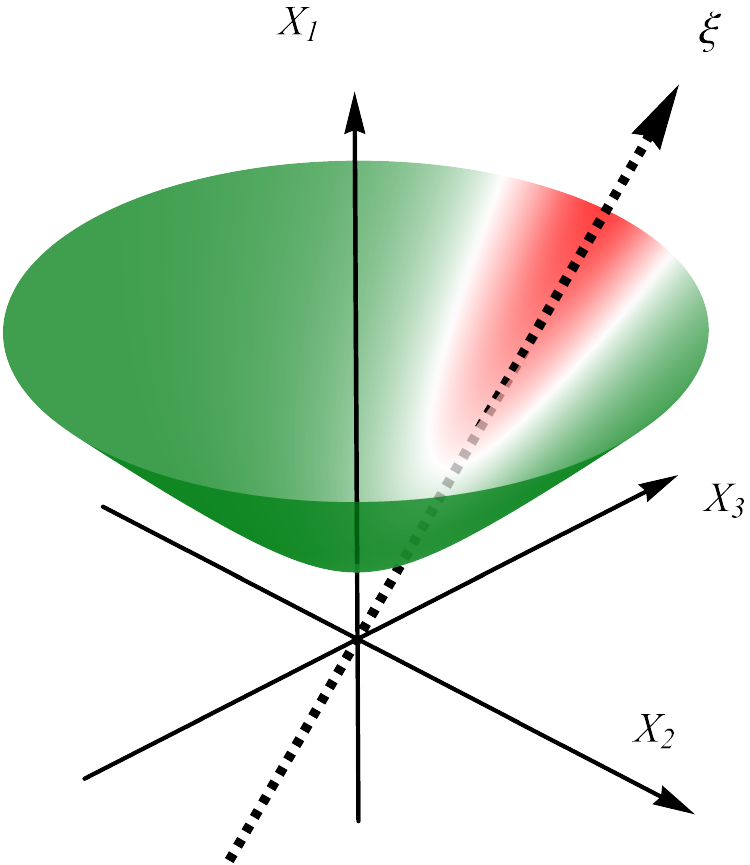}
     \caption{ The value of the $\phi$ field of the soliton in Lobachevsky space for $m=1$ and null vector $\xi=(1,0,1)$. The value of the field changes from $0$ (red) to $2\pi$ (green). }
    \label{lob}
\end{figure}

Let us visualize this solution in two-dimensional Lobachevsky space. In this case the embedding coordinates and null vector are given by: 
\begin{align}
\begin{cases}
X_{0}=\cosh(\psi)
\\X_{1}=\sinh(\psi) \cos(\theta)
\\X_{2}=\sinh(\psi) \sin(\theta)
\end{cases}
\quad \text{and} \quad
\begin{cases}
\xi_{0}=q
\\\xi_{1}=q\cos(\varphi)
\\\xi_{2}=q \sin(\varphi)
\end{cases}.
\end{align}
Then: 
\begin{align}
 -   X\cdot\xi =q\left( \cosh(\psi)-\sinh(\psi) \cos(\theta-\varphi) \right).
\end{align}
Without loss of generality we can choose $\varphi=0$. The graph of this solution is shown in Figure \ref{lob}, where we plot the value of the field on Lobachevsky space in the ambient coordinates; the dotted line shows the direction of the null vector around which the field is localized. 

\section{Conclusion and acknowledgments}

Thus, we have constructed an infinite family of solitonic solutions in $AdS_{d+1}$, $d\geq 2$ spacetime for a deformation of the sine-Gordon theory and for the polynomial potential. In $AdS_{1+1}$, $dS_{d+1}$ and $\mathrm{H}_{d+1}$ for any $d$ we construct a single soliton solution. The crucial point for the presence of multiple soliton solutions is the existence of a two-dimensional null vector space. Due to the absence of such a space in $AdS_{1+1}$, $dS_{d+1}$ and $\mathrm{H}_{d+1}$, we can construct only one solitonic solution. Furthermore, the multiple soliton solutions in $AdS_{d+1}$, $d\geq 2$ reduce in the flat space limit (when such a reduction is possible) to single soliton solutions in flat space.

We hope that this method can be generalized to non-null and non-orthogonal $\eta_i$, $i=\overline{1,N}$ to obtain more general solutions that reduce to the general $n$-soliton sine-Gordon solution in the flat-space limit. However, we cannot exclude the possibility that multiple soliton solutions in hyperbolic spaces cannot exist in principle. The no go theorem can follow from the fact that there are no asymptotic states in such spaces in the proper sense \cite{Akhmedov:2008pu,Akhmedov:2009vh,Akhmedov:2009be}: the existence of multiple soliton solutions that reduce to multiple solitons in flat space would mean that solitons actually represent such asymptotic states of scattering processes. 

Furthermore, sine-Gordon theory in flat space is integrable because of the factorization of the $S$-matrix. But if one cannot define the $S$-matrix \cite{Akhmedov:2008pu,Akhmedov:2009vh,Akhmedov:2009be} in external fields of various nature, then the question is in what sense the theory is integrable in hyperbolic spaces? Should there be some sort of factorization of correlation functions instead of amplitudes?

We would like to thank Bazarov K., Kazarnovski K., Myakutin I., Sadekov D. and Zverev G. for valuable discussions. The work of Diakonov Dmitrii was supported by the grant from the Foundation for the Advancement of Theoretical Physics and Mathematics ``BASIS'', and by the state assignment of the Institute for Information Transmission Problems of the RAS. The work of E.T. Akhmedov was partially supported by the Ministry of Science and Higher Education of the Russian Federation (agreement no. 075–15–2022–287).

\bibliographystyle{unsrturl}
\bibliography{bibliography.bib}

\end{document}